# Effect of nanostructured zinc oxide additives on the humidity and temperature sensing properties of cuprous oxide

B. C. Yadav*[1,2], A. K. Yadav[1] and Anurodh Kumar[1]

[1]*Nanomaterials and Sensors Research Laboratory*
Department of Physics, University of Lucknow, Lucknow-226007, U.P., India

[2]Department of Applied Physics, School for Physical Sciences,
Babasaheb Bhimrao Ambedkar University, Lucknow-226025, U.P., India
*Email: balchandra_yadav@rediffmail.com, nano.lu71@gmail.com
Mob.(+91) 9450094590; Phone. (+91) 5222998125, Fax. (+91) 522 4060185

**Abstract**

Present paper reports the effect of ZnO additives on humidity and temperature sensing properties of cuprous oxide. The cuprous oxide powder was mixed with 10 and 25% ZnO by weight and these samples were pelletized by using hydraulic pressing machine. The sensing materials were also investigated by Scanning Electron Microscope (SEM) and X-ray Diffraction (XRD). SEM images show morphology and porosity of material. The average particles size of cuprous oxide was found to be 1.2 micron. The sheet like structures of ZnO is evident in micrographs. From XRD all peaks are well identified and crystallite size for defferent peaks has also been calculated. The pellets of sensing materials were subjected to annealing at temperatures 200, 400 and 600 ºC respectively and were exposed to humidity and temperature variations. Electrical resistances of pellets were found to vary with humidity and temperature and were recorded. The sensitivity of sensors at various temperature and humidity levels was calculated.

## 1. Introduction

Humidity is defined as the concentration of water molecules in the atmosphere. The measurement of humidity has been proved as a critical task in comparison to other types of environmental parameters like temperature. The influence of humidity has become the main concern for many years in moisture sensitive areas such as high voltage engineering systems, food processing, textile manufacturing, storage areas, hospitals, museums, libraries and geological soil sample studies [1-2]. In particular monitoring moisture content in soil has become a pre-requisite for a variety of processes such as agriculture areas, prone to landslides and laboratory testing application of microsensors in geological and geotechnical engineering study has emerged only recently owing to the complex boundary conditions that must be overcome in granular materials such as three-phase solid-water-air void structure and heterogeneous particle distributions. Recent research in this area has focused on the measurement of suction and humidity in soils [3].

Measurement of moisture content has been guided by the agricultural industry resulting in improved time domain reflectometry devices whose dimensions are far too large for capturing microscale behaviour. Determination of moisture and moisture migration during environmental and physical loading of soils are critical for model development. Direct observation at the microscale of this phenomenon is difficult with the current technology which requires a need for development of microsensors to capture this moisture response.

Various type of moisture sensors are already introduced which are based on different physical sensing principles e.g. resistive [4-6], capacitive [7-8], mechanical [9], gravimetric and thermal humidity sensors [10]. The conductive type sensing principle is based on using porous oxide semiconductors [11-13] such as $MgCr_2O_4$-$TiO_2$, $MnWO_4$ while capacitive type is using polymeric films [14-15]. As far as material for sensors is concerned ceramic humidity sensors show better chemical resistance and mechanical strength with respect to polymeric sensors. The copper (I, II) oxides have a special importance in the application of humidity and temperature sensors.

Ming et al [16] first used the copper (II) oxide as humidity sensor during his studies of thick film. However, there have been a number of works on humidity sensing structures containing copper oxides in their construction such as the CuO/ZnO thin film [17-18]. These investigations indicate that oxides of copper might be useful for fabrication of humidity sensors and thus their properties need further study.

Hence, present paper reports the effect of ZnO additive on the humidity and temperature-sensing characteristics of cuprous oxide.

## 2. Experimental details

### 2.1. Preparation of Sensing Elements

Benedict's reagent was prepared by dissolving 173 gm of sodium citrate and 90 gm of anhydrous sodium carbonate in 500 ml of deionised distilled water. The contents were heated slightly to dissolve. The solution was now filtered and its volume was made to 850 ml. 17.3 gm of $CuSO_4.5H_2O$ was dissolved in 150 ml of distilled water separately. This solution was added slowly with stirring to the above solution, the mixed solution was ready for use. Now a mixture was prepared by adding 50 ml glucose solution (0.2g/ml) to 150 ml Benedict's reagent and then it was boiled for 5 minutes. The proportion was set to give maximum yield of cuprous oxide particles and nearly full consumption of Benedict's reagent and reducing glucose so that we could find highly pure $Cu_2O$. After cooling the boiled mixture, brick red precipitate of cuprous oxide was obtained which was settled down within few minutes. Now filtration of cooled solution gave precipitate of cuprous oxide particles. This precipitate was washed many times by deionised distilled water. The Chemical reactions involved in the process are as following:

$CuSO_4 \rightarrow Cu^{++} + SO_4^{--}$ ……………………. (i)

$Cu^{++}$ + Sodium citrate $\rightarrow$ cupric sodium citrate complex ……………………. (ii)

$$\text{Reducing glucose} \xrightarrow{Na_2CO_3} \text{Enediols forms of reducing glucose} \quad \text{……………..(iii)}$$

Enediols + Cupric: Sodium citrate complex → $Cu^+$ + Mixture of glucose acids ….. (iv)

$$Cu^+ + OH^- \longrightarrow CuOH \quad \text{..............................................(v)}$$

$$2CuOH \xrightarrow{heat} Cu_2O \downarrow + H_2O \quad \text{........................................................ (vi)}$$

The precipitate was dried slowly and brick red fine powder of cuprous oxide was obtained.

ZnO was synthesized through conventional precipitation using hydroxide route "dropwise method" in our laboratory. Detailed characterization had published in international journal of nanotechnology and applications [19]. This ZnO was mixed with cuprous oxide in an amount of 10 and 25% by weight. Pellets of these mixtures with 10% boric acid as binder were made by hydraulic pressing machine [M.B. Instruments, India] under pressure of 615 MPa. Heat treatments of pellets were done at temperatures 200, 400 and 600 °C successively in an electric furnace [Ambassador, India].

## 2.2. Surface Morphology

### 2.2.1. Scanning Electron Micrographs

The surface morphology of sensing materials was investigated by using Scanning Electron Microscope (Philips 505, Netherland). Figs. 1(a) and (b) show the scanning electron micrographs of cuprous oxide with 25% ZnO powder at different magnifications. Sheet like structures of ZnO are evident with small particles of cuprous oxide. Fig. 1(b) shows the micrograph of same sample at higher magnification, in which cuprous oxide particles are observed with average particle size 1.2 micron. Figs. 1(c) and (d) are micrographs of pellet of cuprous oxide annealed at 400 ºC. In Fig. 1(c), the particle structure is not clear but porosity is easily observable. On higher magnification as shown in Fig. 1(d) pores of random structure may be seen.

### 2.2.2. X-ray Diffraction Study

Fig. 2 shows the XRD of cuprous oxide with 25% ZnO annealed at 400 ºC by using X-ray Diffraction (X-Pert PRO system, Netherland). Cuprous oxide starts to transform into cupric oxide at or above 400 ºC in presence of air, therefore peaks of cuprous oxide, cupric oxide and ZnO are found to present in X-ray diffractogram. The highest peak exists at angle $2\theta = 36.5^o$ on the plane (101). This peak corresponds to ZnO. The corresponding values of full width half maxima (FWHM) and 'd' spacing are 0.134º and 2.465 Ǻ respectively. The average size of crystallite to this peak calculated from Scherrer's formula is found to be 653 nm. The minimum size of crystallite is found to be 187 nm for the plane (110) of ZnO at the angle $2\theta = 56.6^o$. The corresponding values of FWHM and 'd' spacing are 0.504º and 1.626 Ǻ respectively. Thus the size of crystallites lies in the range 187 to 1362 nm.

### 2.3. Device Assembly

The prepared sensing elements were exposed to controlled variation of humidity in a self-designed conventional humidity chamber [20] and temperature in a tubular furnace separately. For both type of observations Cu electrode-pellet-Cu electrode conductivity holder was used and variations in resistance were noted with the help of multimeter (Sinometer, VC 9808, India). Relative humidity was measured using standard Hygrometer (Huger, Germany) with least count 1%RH and for temperature measurement an L-shaped laboratory thermometer with least count of 2 °C was used.

## 3. Results and discussion

### 3.1. Humidity Sensor

Variations in resistance with increase in %RH for sensing elements unannealed and annealed at 200 °C have been shown in Figs. 3 and 4. Curve 'a' of Figure 3 represents the variation in resistance of sensing element $Cu_2O$+10% ZnO and shows that resistance of pellet decreases rapidly with increase in %RH. The average sensitivity is found to be 27 MΩ/%RH.

Curve ‘b’ represents the variations in resistance of the sensing element $Cu_2O$+25% ZnO which shows that resistance and hence sensitivity is now increased but measurable range of %RH is reduced due to high resistance. The average sensitivity is now found to be 29 MΩ/%RH. Figure 4 shows that resistance of both sensing elements is decreased significantly after annealing at 200 °C for 4 hrs. Curve ‘a’ indicates that variation in resistance for sensing element with 10% ZnO is almost linear for 10-25% RH and for sensing element with 25% ZnO, curve ‘b’ shows that linearity is up to 50% RH. Beyond these ranges the variation in resistance is slow for both cases. Thus we see that variation in resistance for sensing elements annealed at 200°C are in measurable range and resistances decrease for entire range of %RH. Figs. 5 and 6 show the variation in resistance with increase in %RH for the sensing elements annealed at 400 and 600 °C respectively. Curves ‘a’ and ‘b’ of this figure show that variations in resistance for sensing elements are almost linear with average sensitivity values of 0.02 MΩ/%RH each. Figure 6 shows that after annealing at 600 °C the resistances are slightly increased but there is continuous decrease in resistance for entire range of %RH. The decrease in resistance is sharp up to 50 %RH with average sensitivities 0.285 and 0.375 MΩ/%RH after that decrease is so small that curves appeared to be flat. Detailed discussion of adsorption mechanism of moisture on solid-state humidity sensor is already published in our paper [21].

### 3.2. Temperature Sensor

Variations in resistances with increase in temperature for the sensing elements unannealed and annealed at 200 °C have been plotted in Figs. 7 and 8. Curve ‘a’ of Fig. 7 shows the variations in resistance for sensing element with 10% ZnO which indicates that there is continuous decrease in resistance with increase in temperature for lower range of temperature. As the temperature increases, the decrease in resistance becomes lesser. The average sensitivity for 20-80 °C is found to be 23 MΩ/°C. Curve ‘b’ shows the similar trend of change in resistance for sensing element with 25% ZnO with nearly same average sensitivity. Fig. 8 shows the variation in resistance for sensing elements annealed at 200 °C. Curve ‘a’ represents the change in resistance for sensing element with 10% ZnO which shows that although there is decrease in resistance in the range of temperature 20-80 °C but it is very small and average sensitivity comes out to be about 1 MΩ/ °C. Curve ‘b’ represents the variation in resistance for sensing element with 25% ZnO and indicates that whole curve contains three almost linear segments for ranges 20-24, 24-42 and 42-80 °C with average sensitivities 2.75, 1.7 and 0.4 MΩ/°C. Figs. 9 and 10 describe the observations of variation in resistances of sensing elements annealed at 400 and 600 °C. Curves ‘a’ and ‘b’ of Fig. 9 show that after annealing at 400 °C the resistances with temperature are small and sensitivities are found to be 0.03 and 0.09 MΩ/°C. Curves ‘a’ and ‘b’ of Figure 10 show that for sensing elements with 10 and 25% ZnO the variation in resistances are similar to that in case of pure cuprous oxide. This is expected result and can be understood on the basis of calcination temperature of ZnO. Since calcination temperature for ZnO is 450 °C so after calcining above this temperature, it should behave like a pellet of pure cuprous oxide. The average sensitivities for sensing elements annealed at 600°C is found to be 0.4 and 0.5 MΩ/°C.

**1.3. Hysteresis Study**

The hysteresis study was performed for sensing element with 10% ZnO having good sensitivity and range of measurement. The data for hysteric behaviour of humidity sensor for sensing element with 10% ZnO is shown in Fig. 11. The average hysteresis is found to be 41%. Fig. 12 shows the hysteric behaviour of temperature sensor. The average hysteresis in this case is found to be 17%.

**3.4. Study of Ageing Effect**

The ageing effects for humidity and temperature sensing behaviour of cuprous oxide with 10% ZnO were also investigated. For this study samples were kept in laboratory environment and observations were taken at the interval of approximately 8 months. Fig. 13 shows the ageing effect of humidity sensor. It is found that sensor became less sensitive. It was deteriorated by 18%. Similarly Fig. 14 shows the ageing effect for temperature sensor. It is found that sensitivity of sensor has decreased by 10%.

**4. Conclusions**

The electrical resistance values of the cuprous oxide pellets with different compositions of ZnO were found to be increased. The sensitivity of humidity sensor was also increased with the amount of ZnO added but resistance became quite high so measurable range of humidity decreased. Thus for higher range of humidity cuprous oxide mixed with ZnO may be a good sensing material. Hysteresis was also decreased when ZnO was added to cuprous oxide. For entire range of humidity measurement, cuprous oxide with 25% ZnO annealed at 200 °C for 4 hrs was found to be a good sensing material as in this case resistance became smaller and measurable but sensitivity is now reduced to 12 MΩ/%RH. In case of temperature sensor, unannealed cuprous oxide with 10% ZnO was found to be the best sensing material having good sensitivity i.e. 23 MΩ/°C and steady variation of resistance.

## Acknowledgement

We are highly grateful to Dr. C. D. Dwivedi, Retd. Scientist, DMSRDE, Kanpur for fruitful discussion on manuscript and corresponding author is thankful to Department of Science and Technology, India for financial support in the form of FAST-TRACK project (SR/FTP/PS-21/2009).

**Figures Captions**

**Fig. 1(a).** SEM of $Cu_2O$ + 25% ZnO powderwithout annealing at agnification 1.62 KX.

**Fig. 1(b).** SEM of $Cu_2O$ + 25% ZnO powder without annealing at magnification 6.50 KX.

**Fig. 1(c).** SEM of $Cu_2O$ + 25% ZnO pellet annealed at 400ºC at magnification 6.50 KX

**Fig.1(d).** SEM of $Cu_2O$ + 25% ZnO pellet annealed at 400ºC at magnification 10.0 KX

**Fig. 2.** X-ray Diffraction pattern of sensing material $Cu_2O$ + 25% ZnO annealed at 400 °C.

**Fig. 3.** Variations in Resistance in MΩ against RH% for sensing elements $Cu_2O$ + 10% ZnO and $Cu_2O$ + 25% ZnO before annealing.

**Fig. 4.** Variations in Resistance in MΩ against RH% for sensing elements $Cu_2O$ + 10% ZnO and $Cu_2O$ + 25% ZnO before annealed at 200ºC.

**Fig. 5.** Variations in Resistance in MΩ against RH% for sensing elements $Cu_2O$ + 10% ZnO and $Cu_2O$ + 25% ZnO annealed at 400 ºC.

**Fig. 6** Variations in Resistance in MΩ against RH% for sensing elements $Cu_2O$ + 10% ZnO and $Cu_2O$ + 25% ZnO annealed at 600 ºC.

**Fig. 7.** Variations in Resistance in MΩ against temperature for sensing elements $Cu_2O$ + 10% ZnO and $Cu_2O$ + 25% ZnO before annealing.

**Fig. 8** Variations in Resistance in MΩ against temperature for sensing elements $Cu_2O$ + 10% ZnO and $Cu_2O$ + 25% ZnO annealed at 200ºC

**Fig. 9** Variations in Resistance in MΩ against temperature for sensing elements $Cu_2O$ + 10% ZnO and $Cu_2O$ + 25% ZnO annealed at 400 ºC.

**Fig. 10.** Variations in Resistance in MΩ against temperature for sensing elements $Cu_2O$ + 10% ZnO and $Cu_2O$ + 25% ZnO annealed at 600 ºC.

**Fig. 11** Hysteresis curve 'Resistance Vs RH%' for the sensing element unannealed $Cu_2O$ +10% ZnO before annealing.

**Fig.12** Hysteresis curve 'Resistance Vs Temperature' for the sensing element unannealed $Cu_2O$ +10% ZnO before annealing.

**Fig. 13** Variation in resistance with RH% at the time interval of 8 months for the sensing element $Cu_2O$ + 10% ZnO annealed at 600 °C.

**Fig. 14** Variation in resistance with temperature at the time interval of 8 months for the sensing element $Cu_2O$ + 10% ZnO annealed at 600 °C.

## Figures

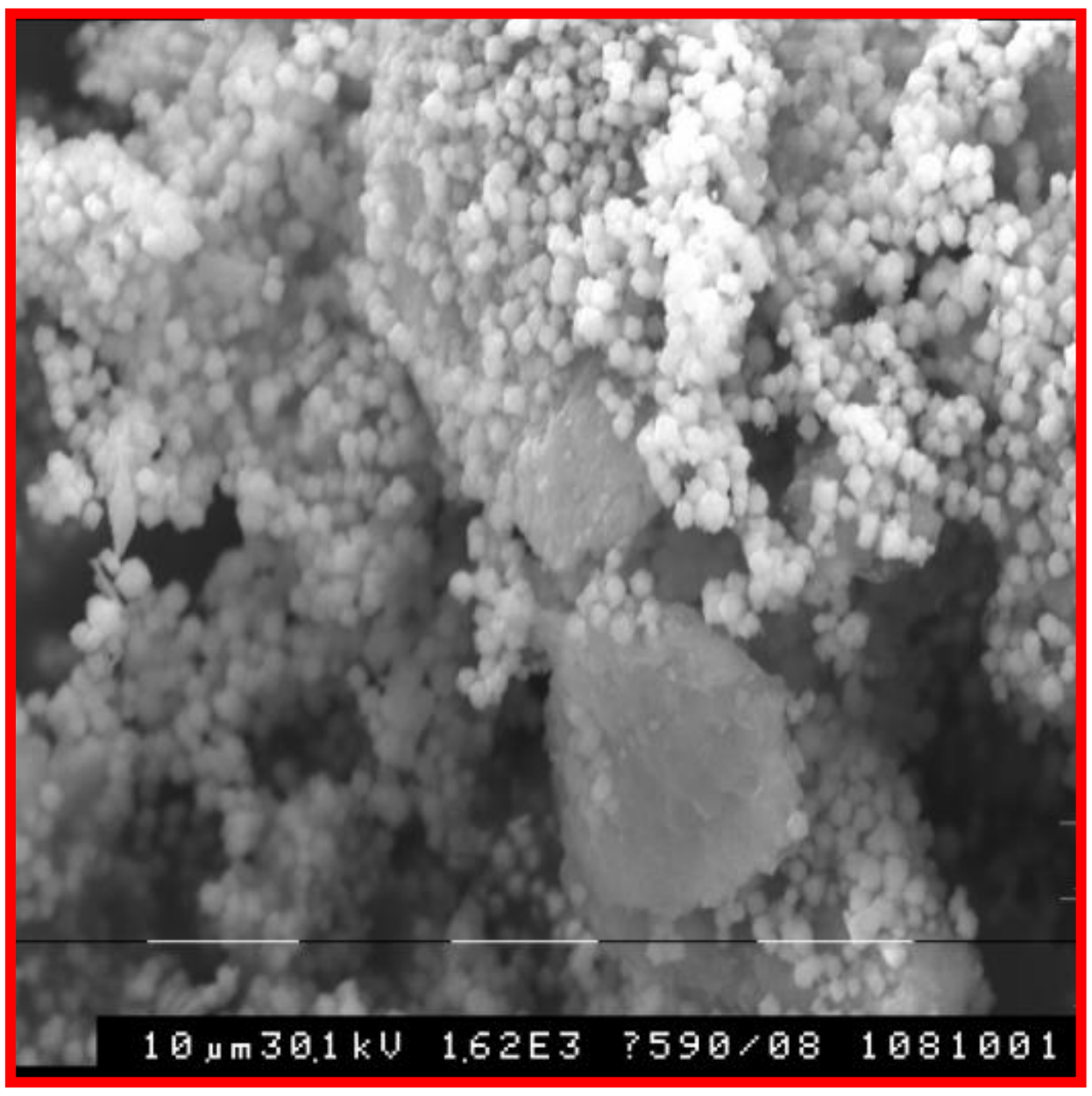

**Fig. 1(a)**

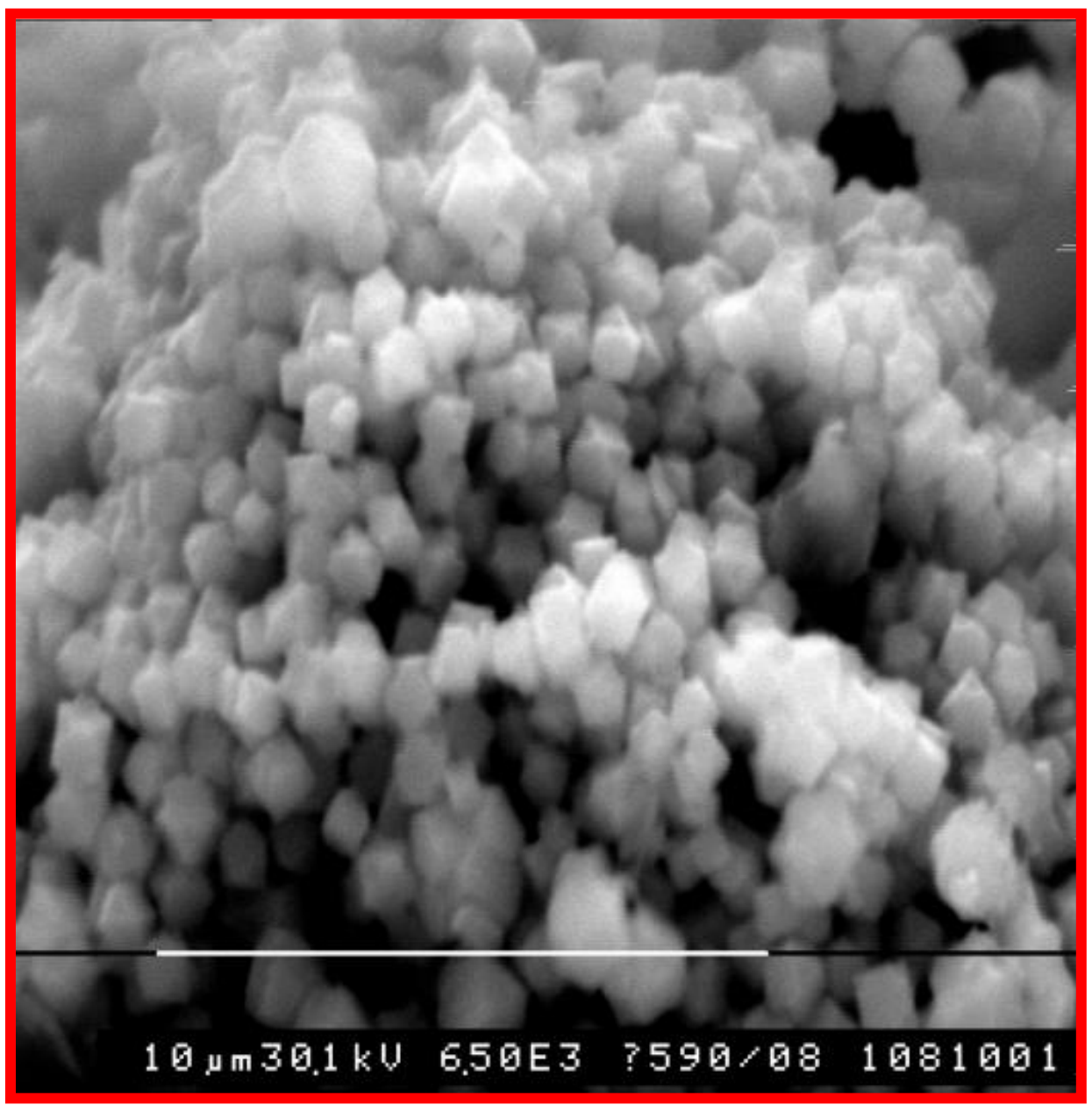

**Fig. 1(b)**

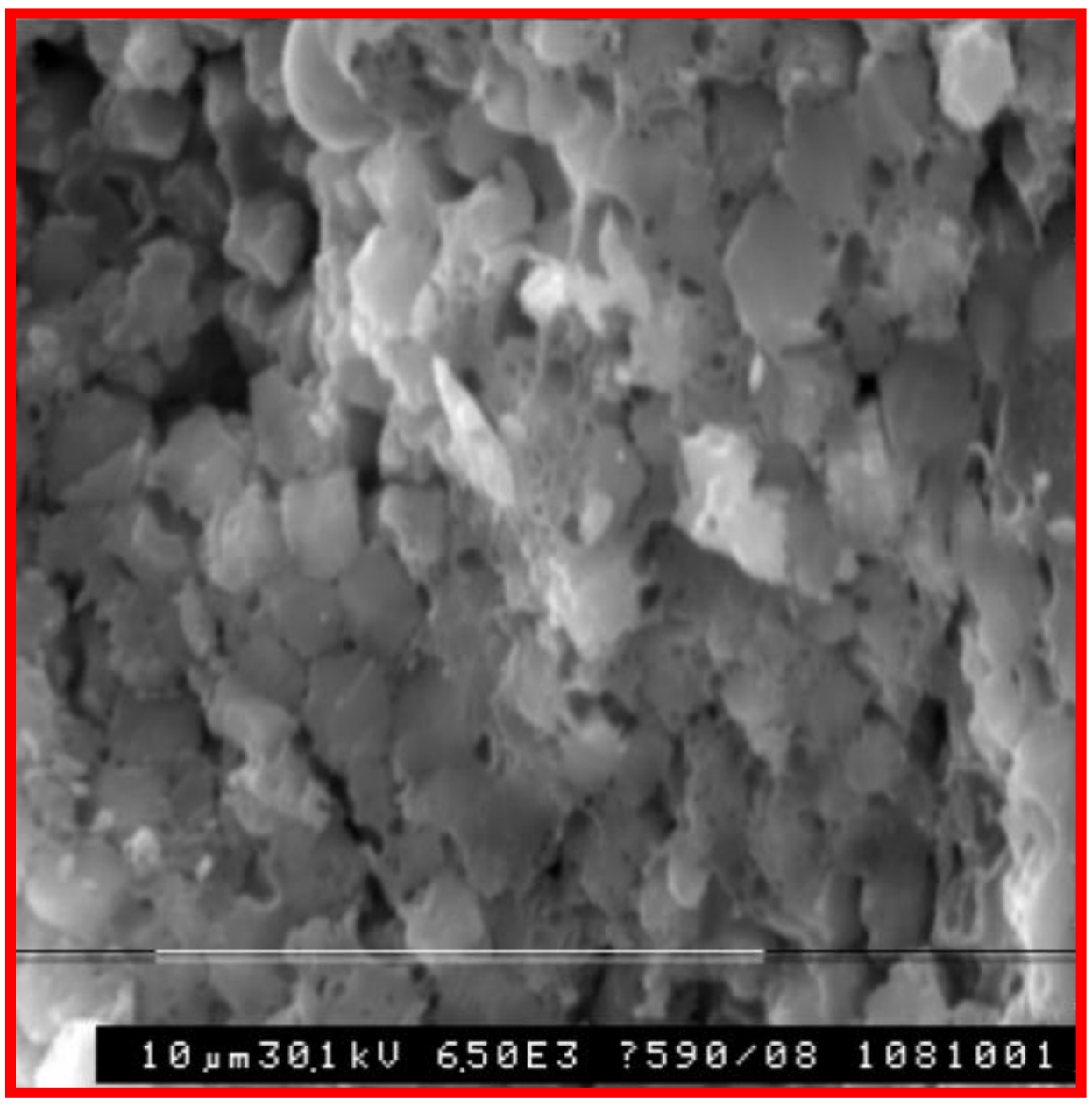
10 µm 30.1 kV 6.50E3 7590/08 1081001

**Fig. 1(c)**

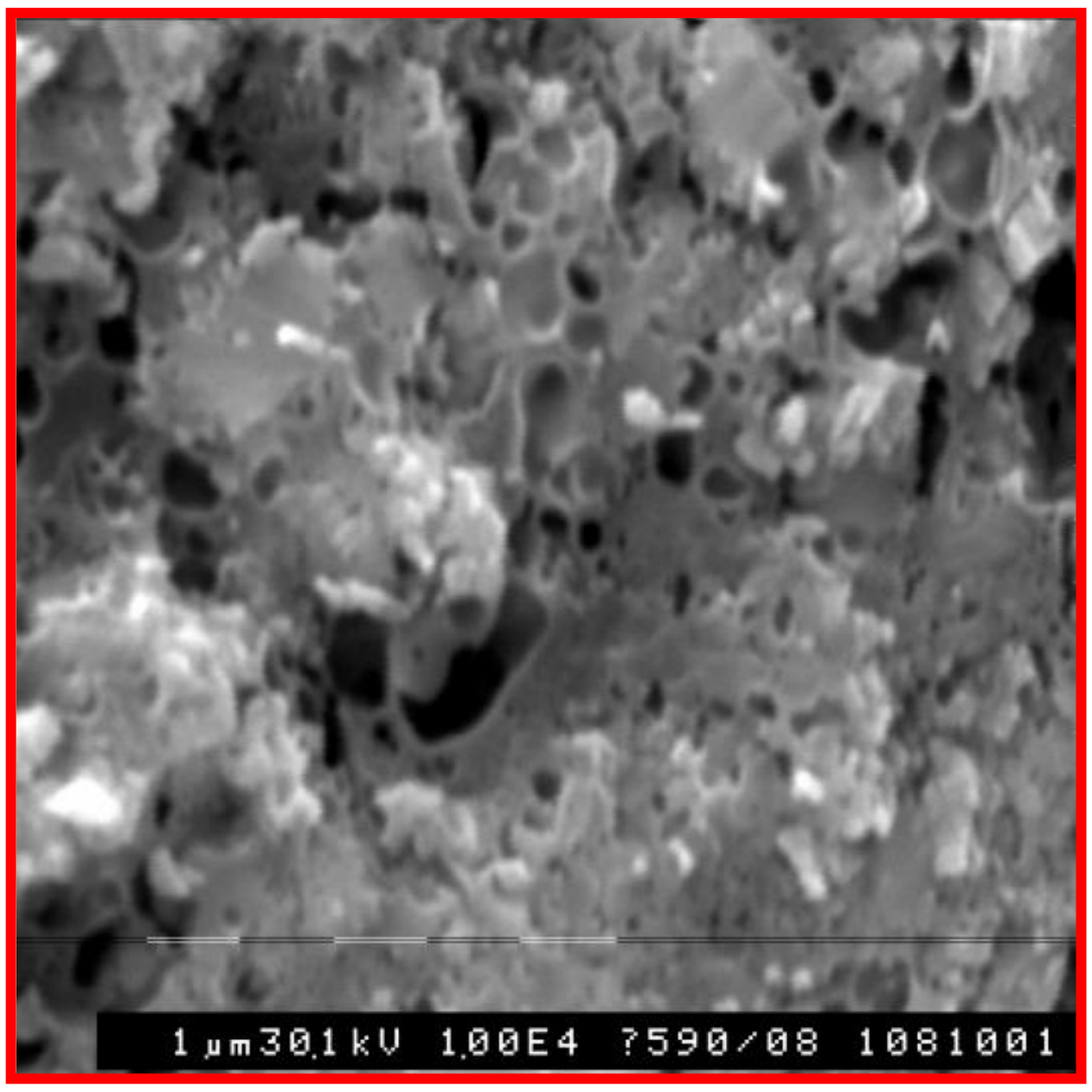
1 µm 30.1 kV 1.00E4 7590/08 1081001

**Fig. 1(d)**

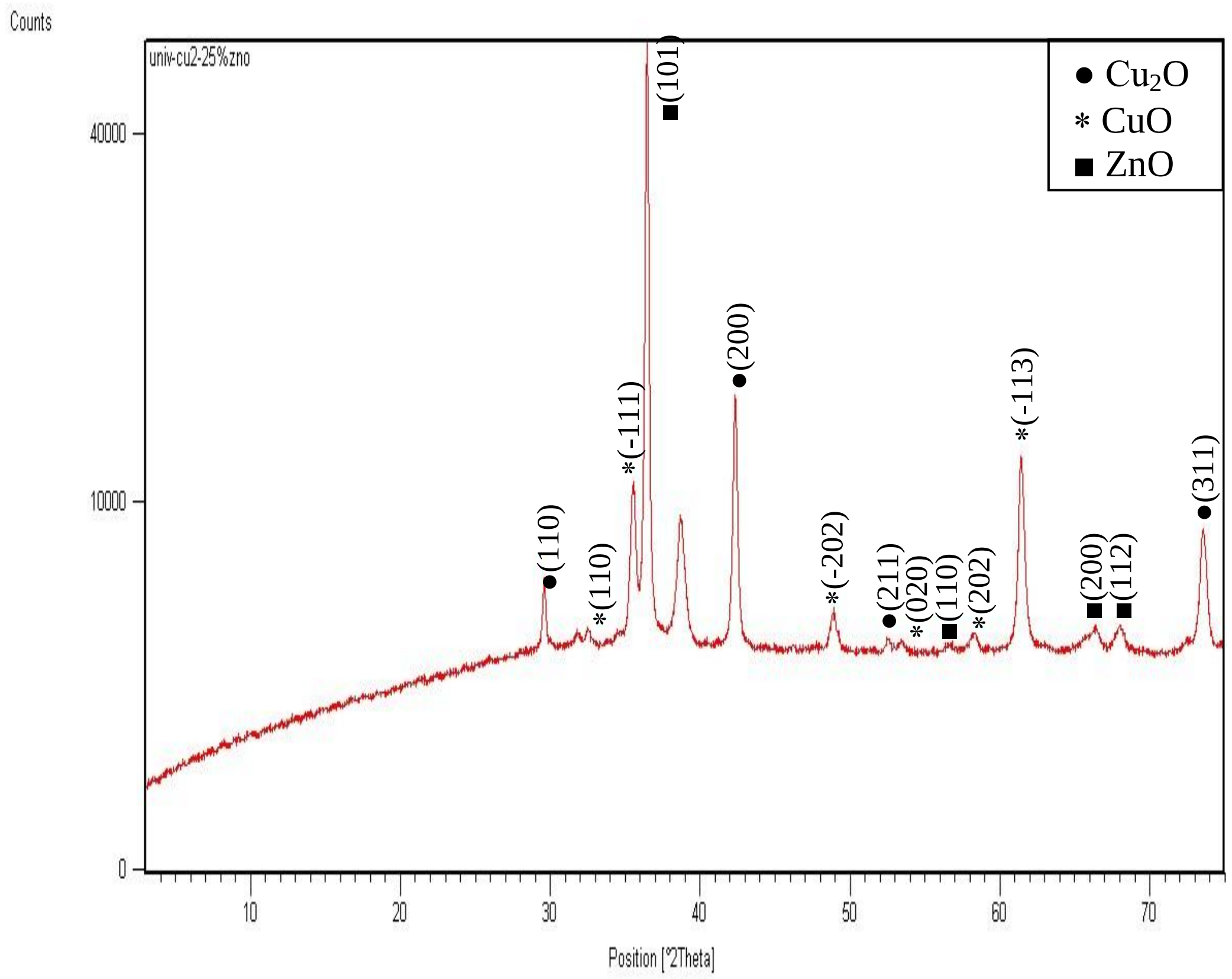

**Fig. 2**

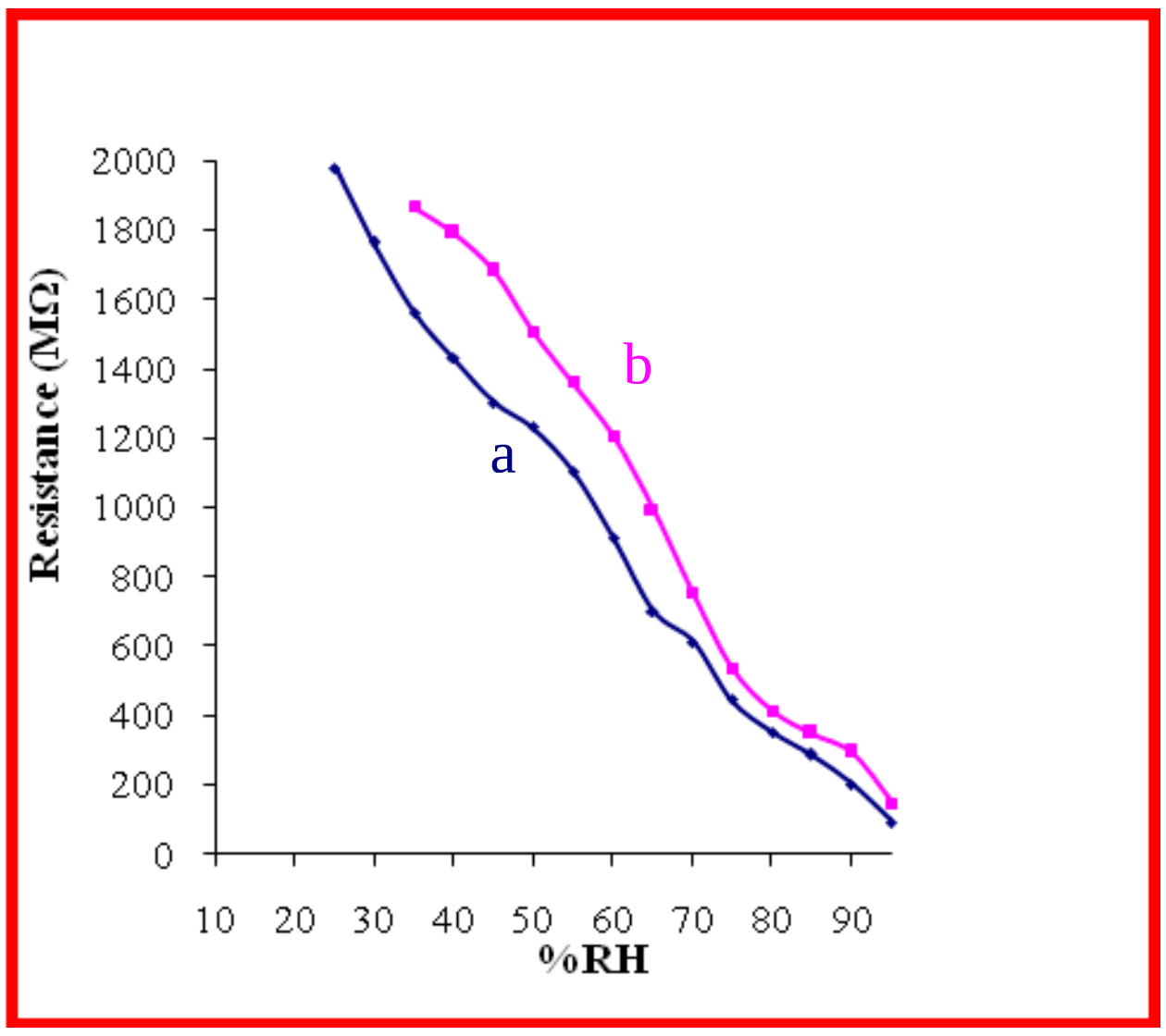

2000
1800
1600
1400
1200
1000
800
600
400
200
0
Resistance (MΩ)
10 20 30 40 50 60 70 80 90
%RH
a
b

**Fig. 3**

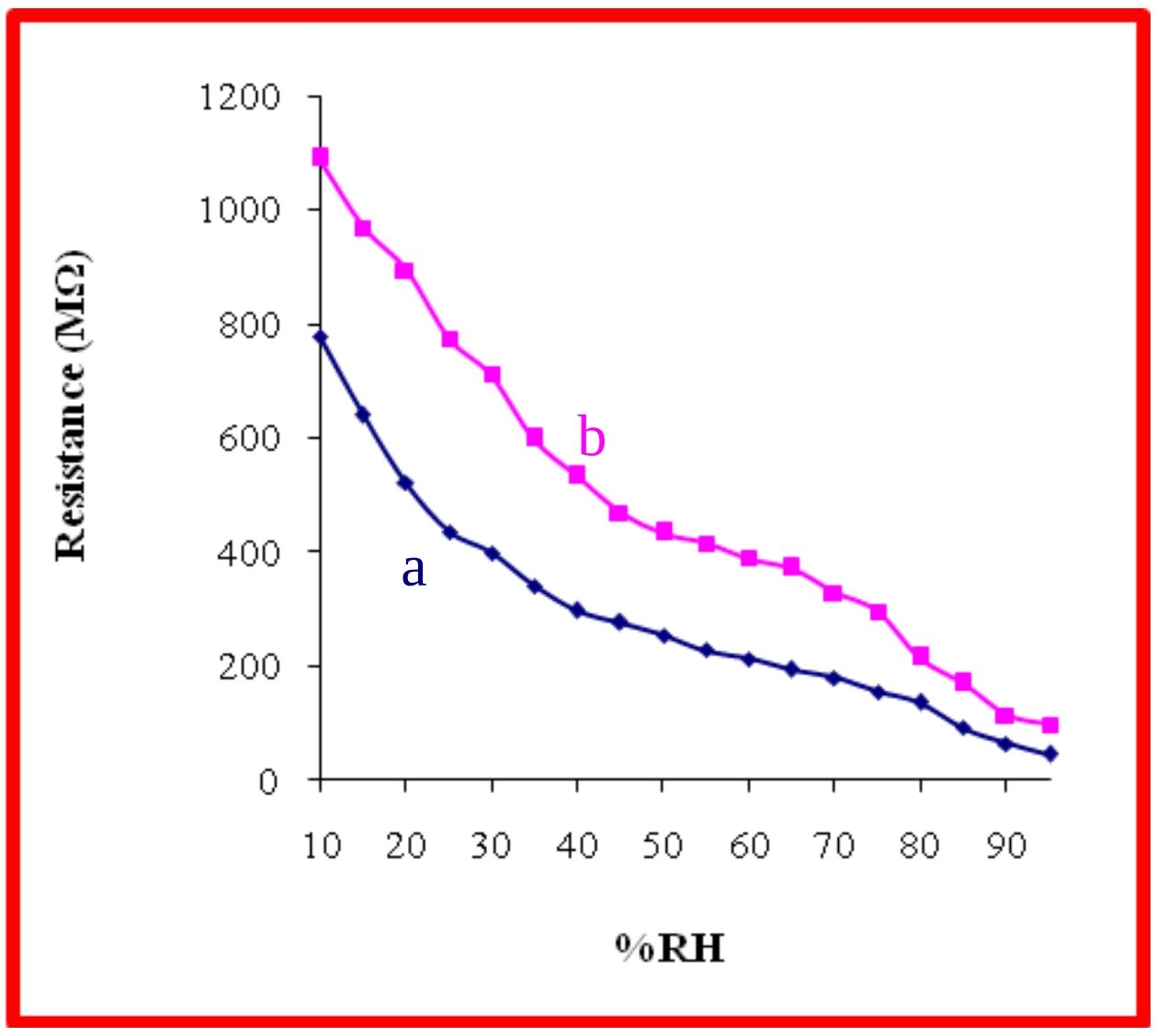

**Fig. 4**

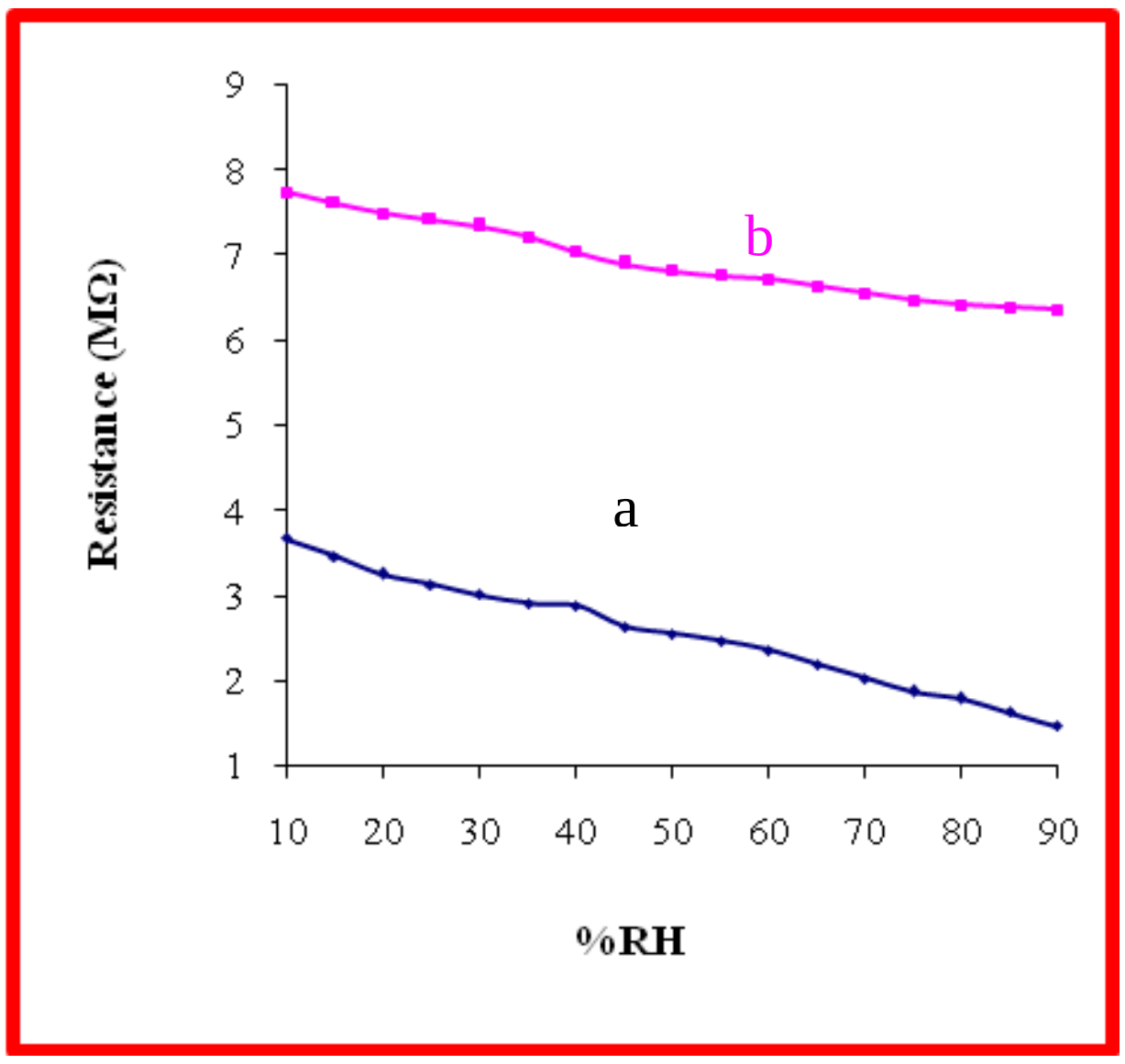

**Fig. 5**

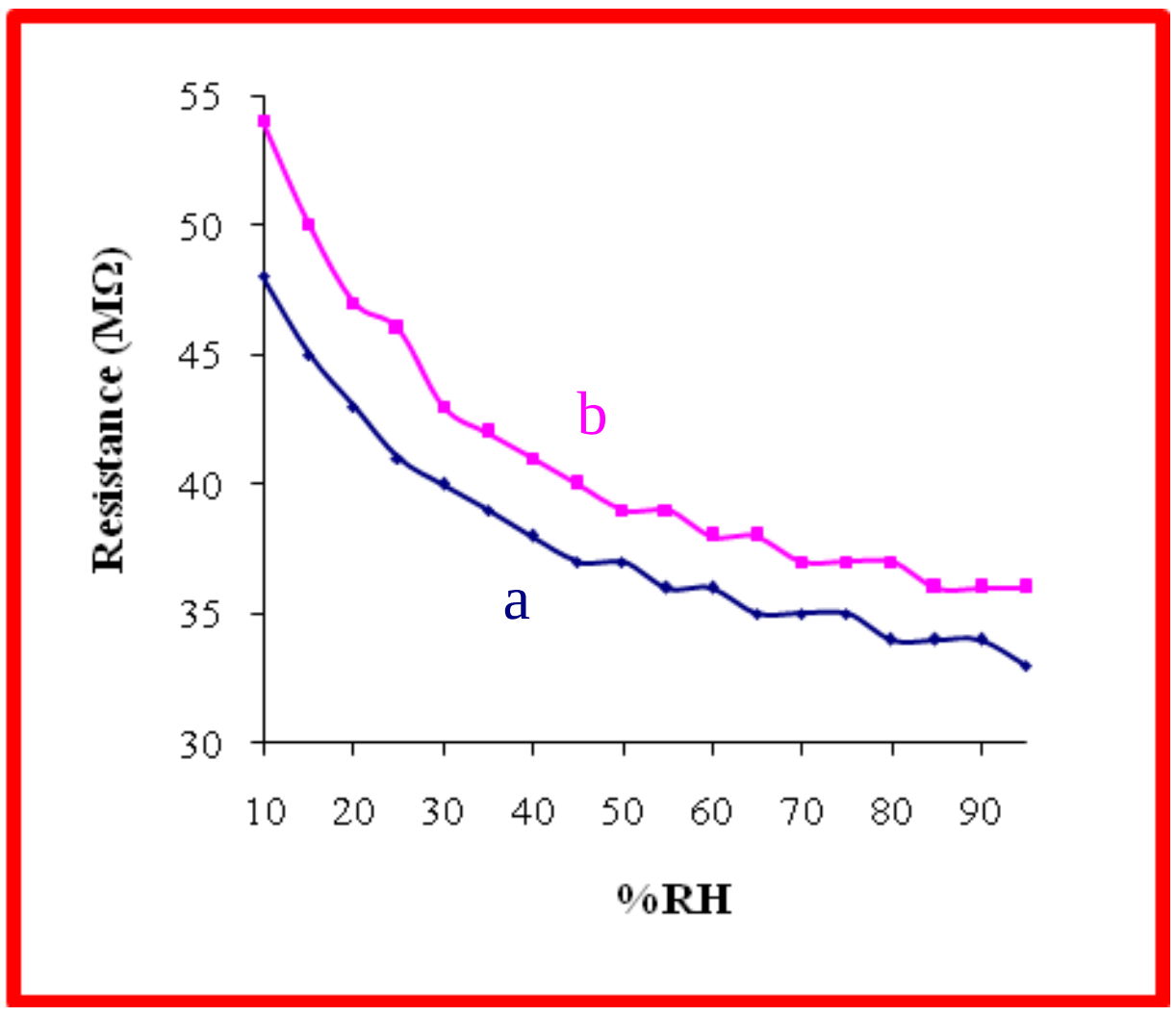
55
50
45
40
35
30
Resistance (MΩ)
b
a
10 20 30 40 50 60 70 80 90
%RH

**Fig. 6**

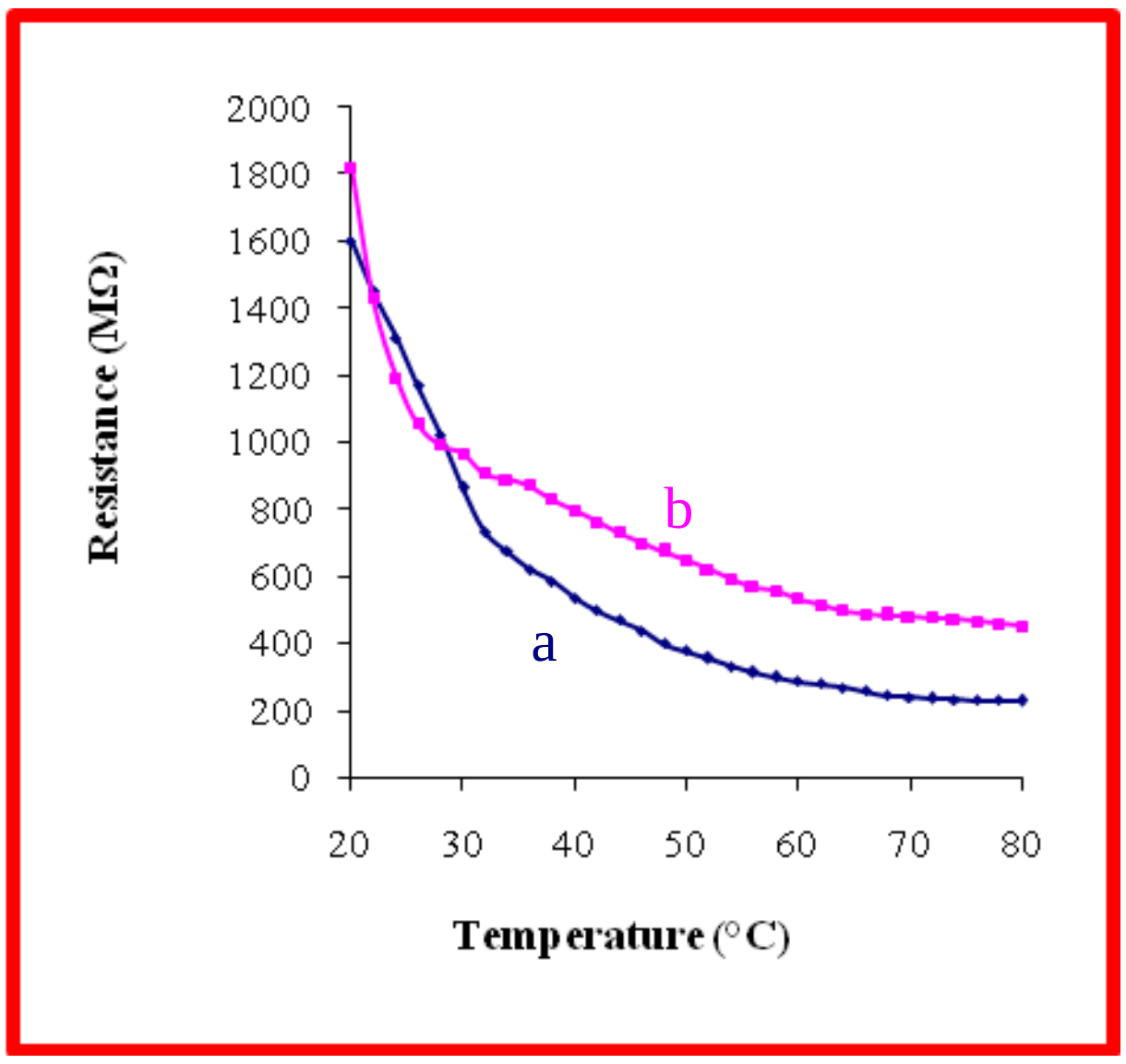

**Fig. 7**

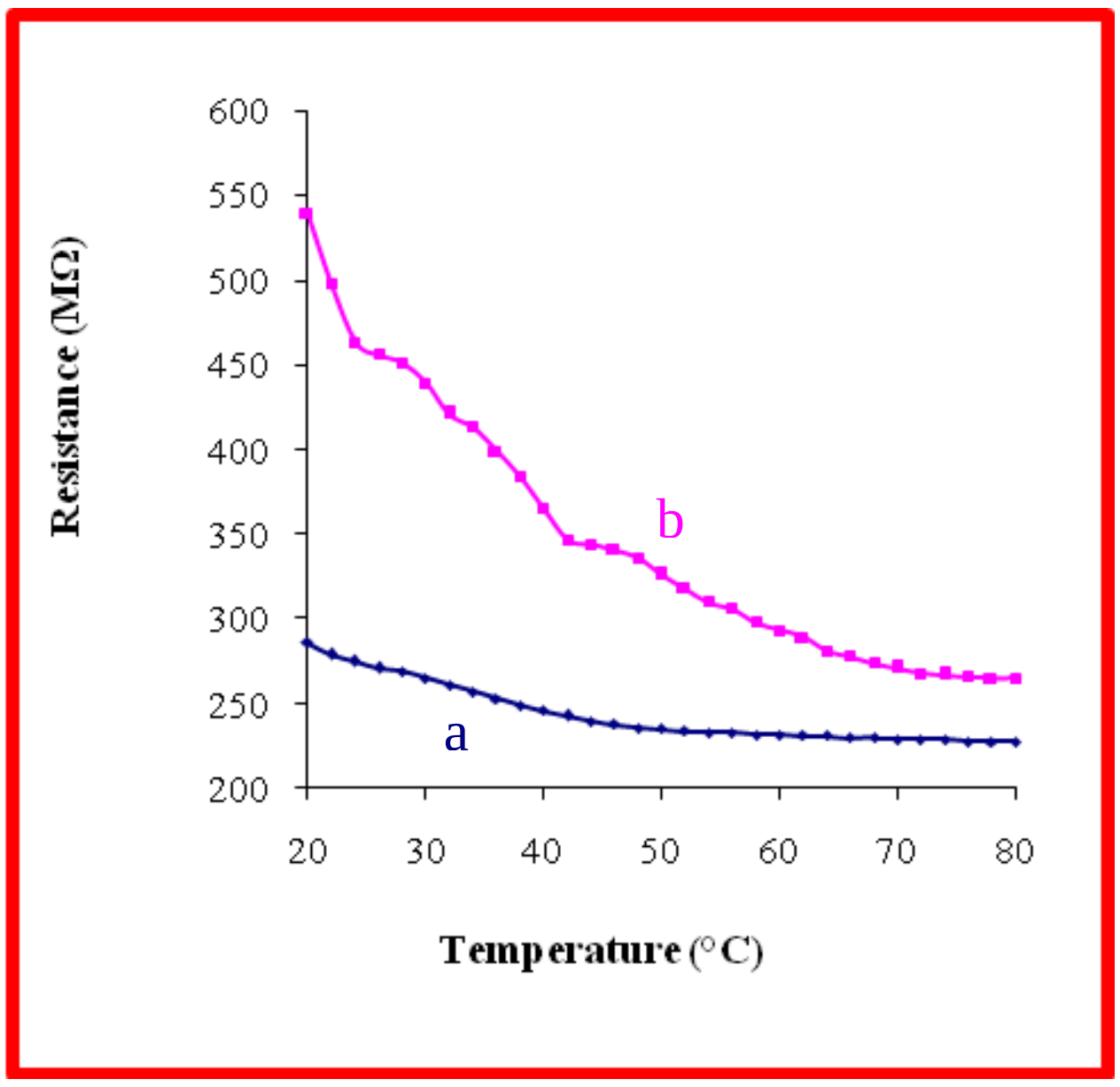

**Fig. 8**

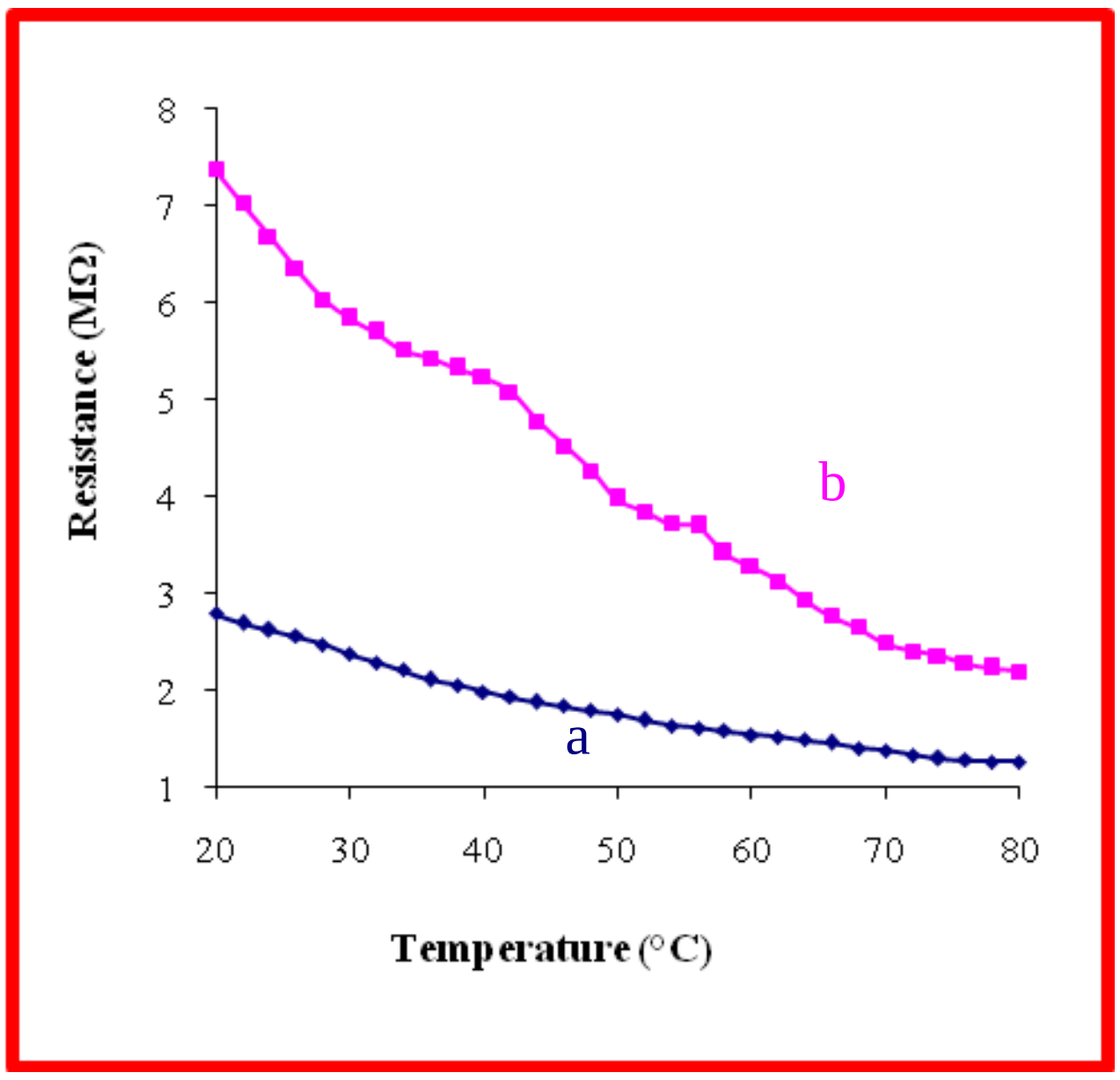

**Fig. 9**

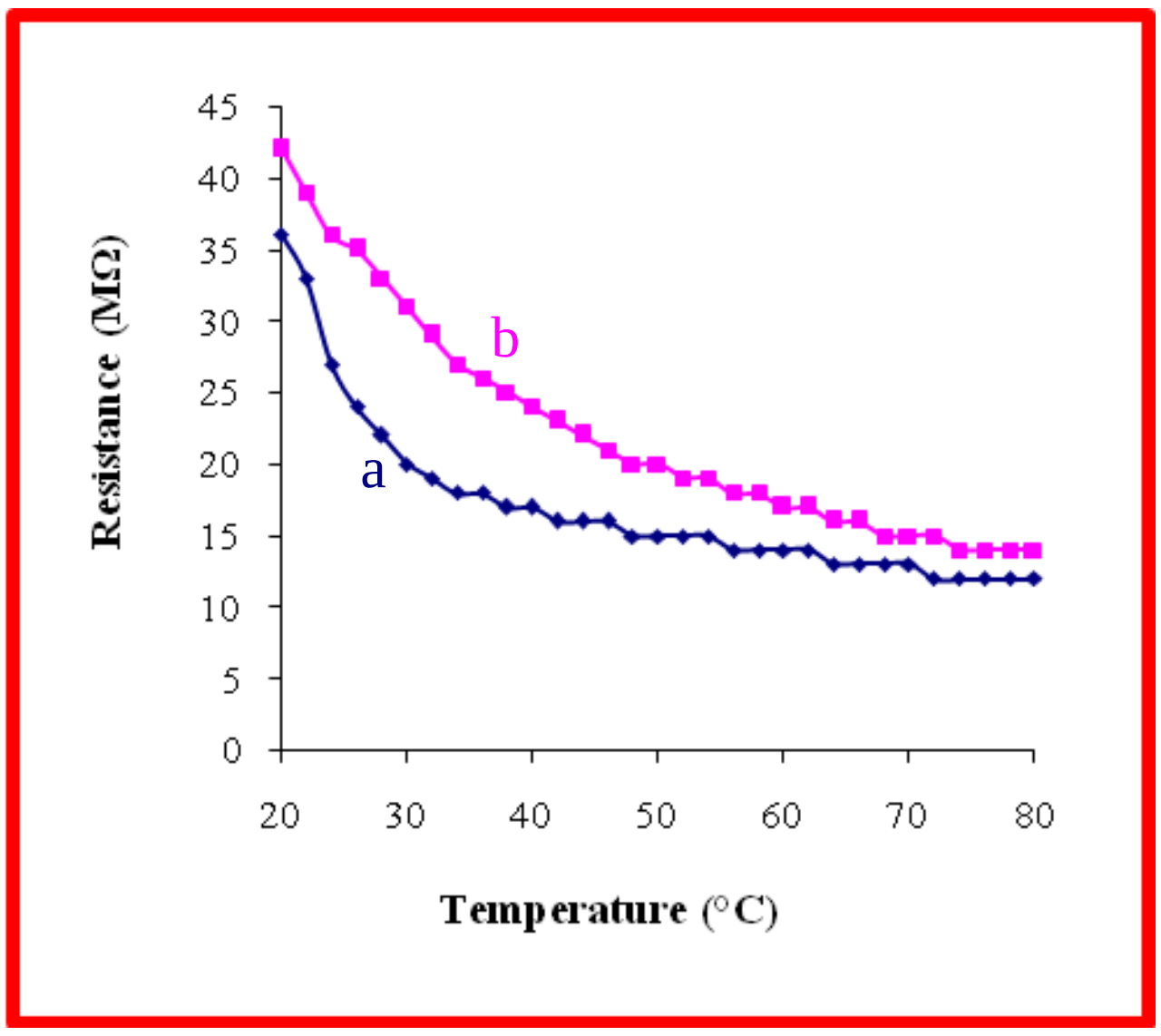

**Fig. 10**

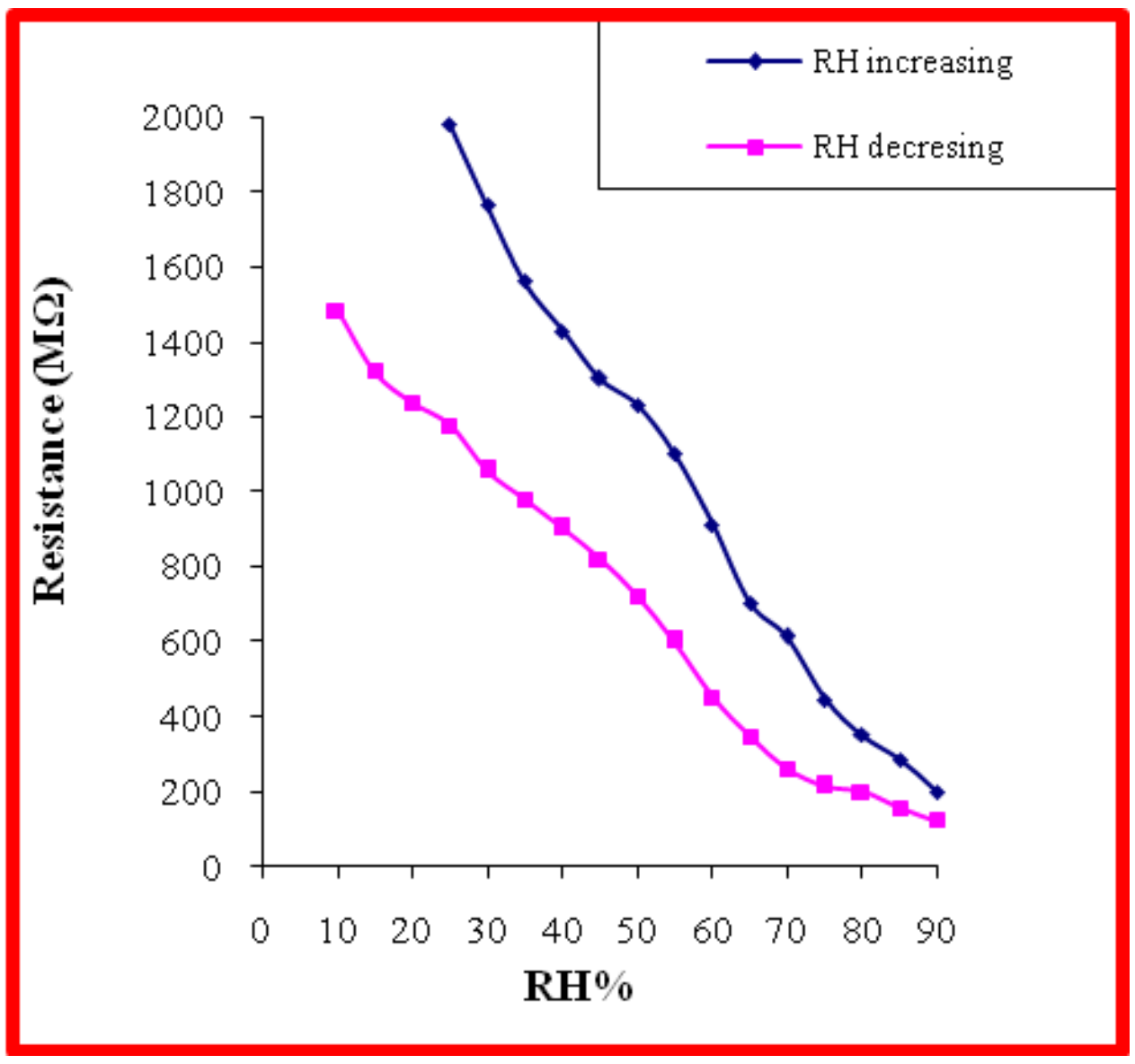

**Fig.11**

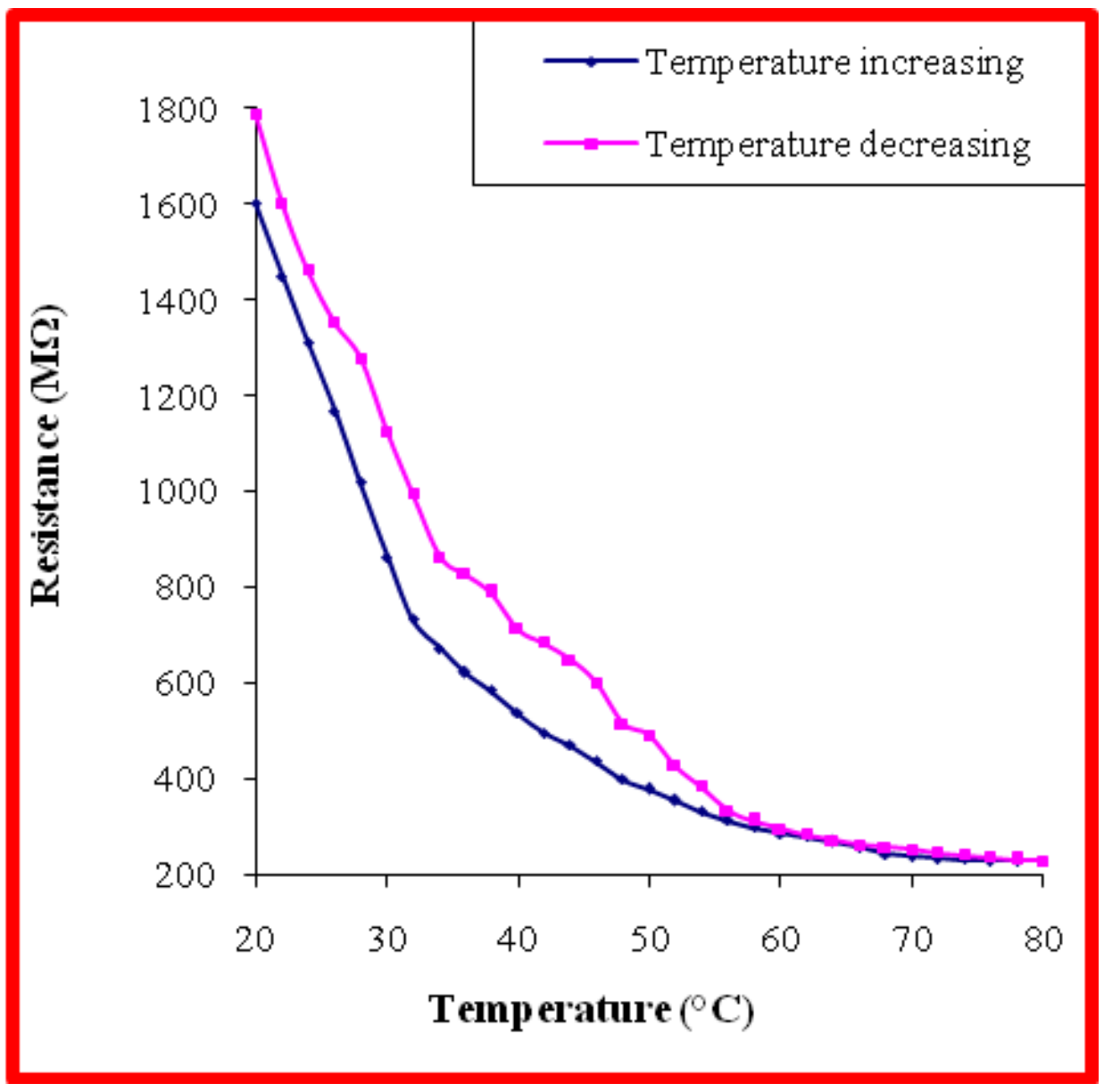

**Fig. 12**

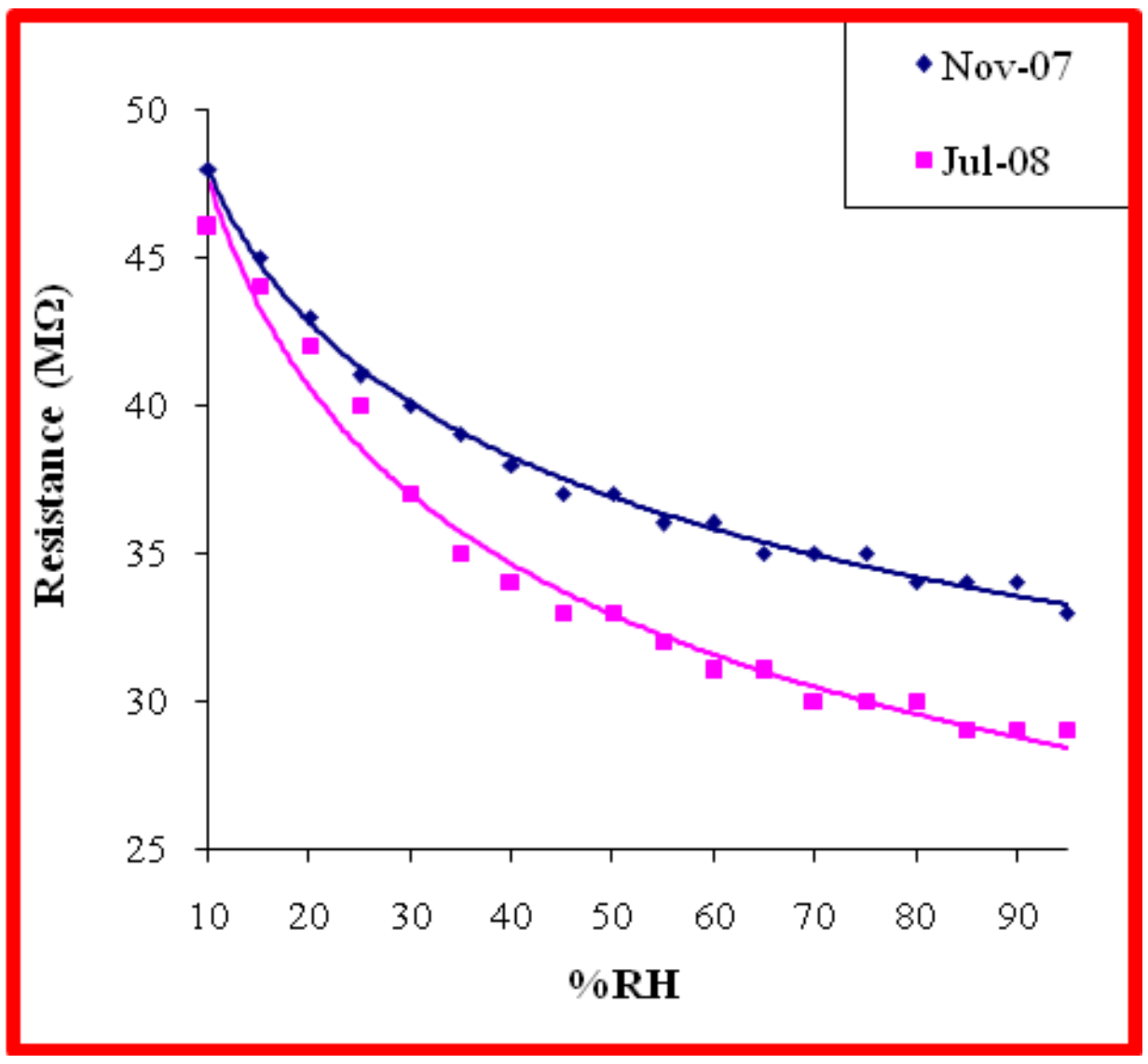

**Fig. 13**

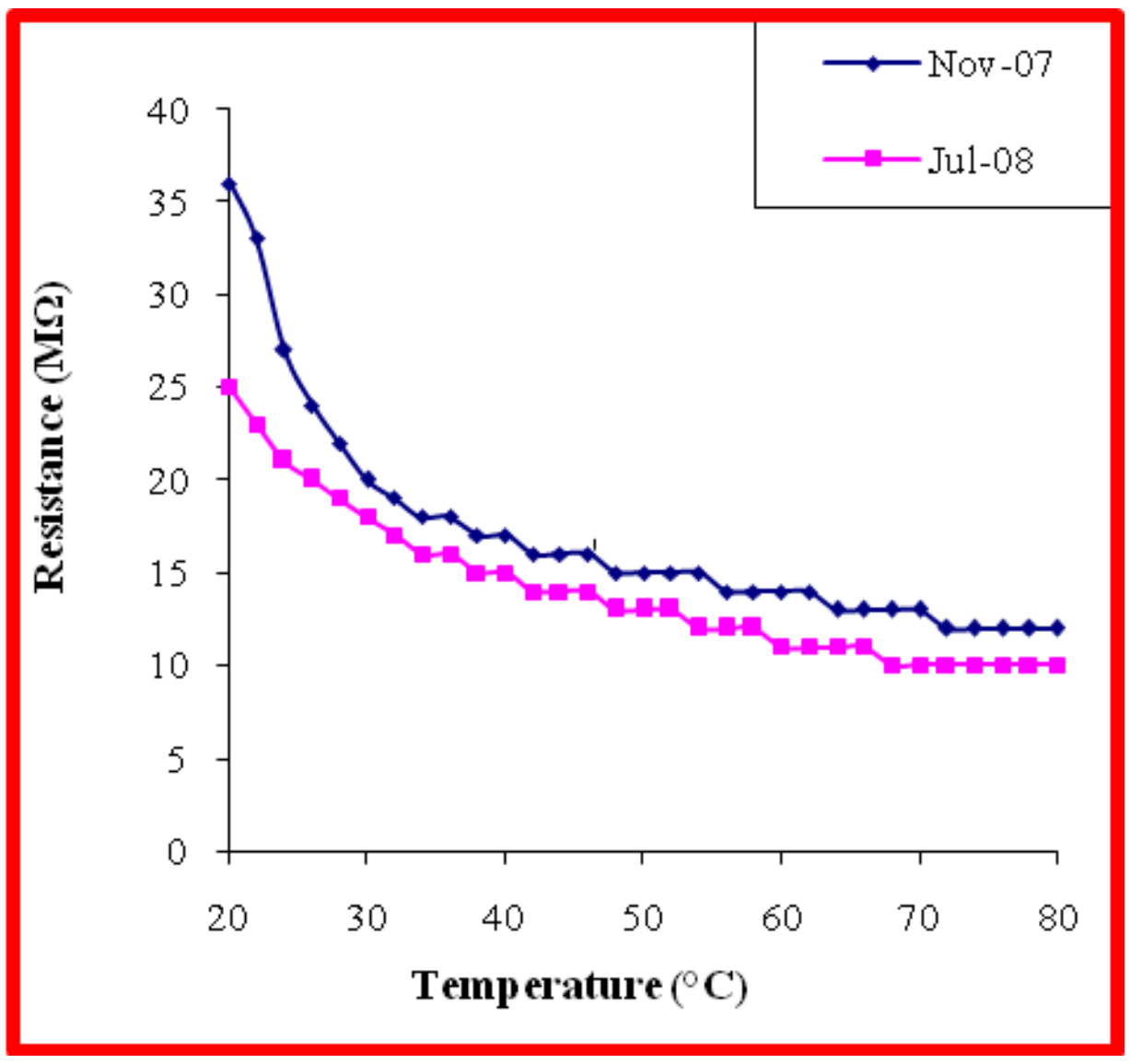

**Fig. 14**